\documentclass{emulateapj}
\RequirePackage{lineno}
\usepackage{amssymb}
\usepackage{amsmath}
\usepackage{mathrsfs}
\usepackage{natbib}
\usepackage[colorlinks, linkcolor=blue, urlcolor=blue, citecolor=blue]{hyperref}
\usepackage{array}
\usepackage{float}
\usepackage{graphicx}
\usepackage{subfigure}
\usepackage{color}
\usepackage[all]{hypcap}
\usepackage[toc,page]{appendix}
\usepackage{diagbox}
\usepackage{booktabs}
\usepackage{multirow}

\def\beq{\begin{equation}}
\def\enq{\end{equation}}
\def\bea{\begin{eqnarray}}
\def\ena{\end{eqnarray}}

\begin{document}

\title{Double-peaked Pulse Profile of FRB 200428: Synchrotron Maser Emission from Magnetized Shocks Encountering a Density Jump}
\author{Di Xiao\altaffilmark{1,2}, and Zi-Gao Dai\altaffilmark{1,2}}
\affil{\altaffilmark{1}School of Astronomy and Space Science, Nanjing University, Nanjing 210093, China; dxiao@nju.edu.cn; dzg@nju.edu.cn}
\affil{\altaffilmark{2}Key Laboratory of Modern Astronomy and Astrophysics (Nanjing University), Ministry of Education, China}

\begin{abstract}
Very recently a fast radio burst (FRB) 200428 associated with a strong X-ray burst from the Galactic magnetar SGR 1935+2154 has been detected, which is direct evidence supporting the magnetar progenitor models of FRBs. Assuming the FRB radiation mechanism is synchrotron maser emission from magnetized shocks, we develop a specific scenario by introducing a density jump structure of upstream medium, and thus the double-peaked character of FRB 200428 is a natural outcome. The luminosity and emission frequency of two pulses can be well explained in this scenario. Furthermore, we find that the synchrotron emission of shock-accelerated electrons is in the X-ray band, which therefore can be responsible for at least a portion of observed X-ray fluence. With proper upgrade, this density jump scenario can be potentially applied to FRBs with multiple peaks in the future.
\end{abstract}


\section{Introduction}
\label{sec1}
Fast radio bursts are bright millisecond radio pulses that flash randomly in the universe. The first discovery was reported in 2007 \citep{Lorimer2007} and there are hundreds of FRBs collected now \citep{Petroff2016}. This field is still at an early stage and many aspects of FRBs remain mysterious, especially the origin of them \citep[for reviews, see][]{Katz2018, Popov2018, Petroff2019,Cordes2019}. Tens of progenitor models have been proposed (see \citet{Platts2019} for a theory catalogue), most of which involve compact objects according to the energetics and temporal variability of FRBs. In recent years, a few observational breakthroughs have been made, which greatly help us understand this phenomenon. The localization of ten FRBs has been realized with their host galaxies being identified \citep{Tendulkar2017, Bannister2019, Prochaska2019, Ravi2019, Chittidi2020, Marcote2020, Macquart2020}. The fact that some FRBs lie in the outskirt of host galaxies potentially points to neutron stars (NS). The periodical activity of two FRBs have been found \citep{CHIME2020a,Rajwade2020b, Cruces2020}, which also seems to be in favor of NS progenitors. Moreover, up to now about twenty FRBs have been found to repeat \citep{CHIME2019a,CHIME2019b,Fonseca2020}, implying that sustainable energy reservoirs are required. It has been suggested that young flaring magnetars could play this role \citep{Popov2013,Lyubarsky2014,Murase2016,Beloborodov2017,Waxman2017}.

The above conjecture was supported early in this year, as a special FRB 200428 from the Galactic magnetar SGR 1935+2154 has been detected \citep{Bochenek2020,CHIME2020b}. Nearly the same time, an associated X-ray burst has been observed by {\em Insight}-HXMT, INTEGRAL, Konus-Wind and AGILE \citep{Li2020,Mereghetti2020,Ridnaia2020,Tavani2020}. This landmark discovery directly verified that magnetars are progenitors of at least a portion of FRBs. SGR 1935+2154 is a typical magnetar with spin period $\sim3.24\,\rm s$ and surface magnetic field strength $\sim2.2\times10^{14}\,\rm G$ \citep{Israel2016}. Several measurements of its distance gave different values, within the range of $4.4-12.5\,\rm kpc$ \citep{Kothes2018,Zhou2020,Zhong2020,Mereghetti2020}. FRB 200428 has two narrow pulses separated by $\Delta T=28.91\,\rm ms$. The first pulse was detected only by CHIME and the second was detected simultaneously by CHIME and STARE2. Moreover, the first pulse cuts off quickly beyond 550 MHz while the second pulse spans from 500 MHz up to 1468 MHz. The combined total fluence of second pulse is at least ten times higher than the first one. All these properties need to be addressed properly in a detailed theoretical model.

The radiation mechanism of FRBs is still largely unknown. The high brightness temperature implies that the radiation must be coherent. Generally there are three kinds of astrophysical approaches to generate coherence: coherent curvature emission by bunches, plasma emission by reactive instabilities and maser emission \citep[for a review, see][]{Melrose2017}. In this work, we mainly consider the synchrotron maser mechanism by magnetized shocks, which was first proposed in the early 1990s. Bunching in gyration phase of incoming particles near shock front has been verified by Particle-in-Cell (PIC) simulations \citep{Hoshino1991,Hoshino1992,Amato2006,Plotnikov2019}. The distribution of the particles in momentum space can be described as a cold ring form \citep{Alsop1988}, which is beneficial to the development of synchrotron maser instability. The growth of electromagnetic (EM) waves can be very effective and thus a maser (FRB) can be produced \citep{Hoshino1991}. In this scenario, physical condition of shock upstream medium could have a major impact on observed FRB properties. For instance, \citet{Metzger2019} and \citet{Beloborodov2020} assumed slow ion tail of flare ejecta and magnetar wind as upstream medium respectively, therefore their predictions are different in many ways. This motivates us to study the influence on FRB emission if a density structure of external medium exists. More intriguingly, recently \citet{Kirsten2020} reported a repetition of FRB 200428 with double-peaked character again after long time monitoring, which supports our hypothesis of a stratified medium.

This letter is organized as follows. We introduce the density jump scenario in Section \ref{sec2}. In Section \ref{sec3} we discuss the dynamics and constrain the density properties from the radio observation. Then the post-shock high-energy emission is discussed in Section \ref{sec4}. We finish with discussion and conclusions in Section \ref{sec5}.

\section{Density jump Scenario}
\label{sec2}
\begin{figure}[H]
\label{fig1}
\begin{center}
\includegraphics[width=0.42\textwidth]{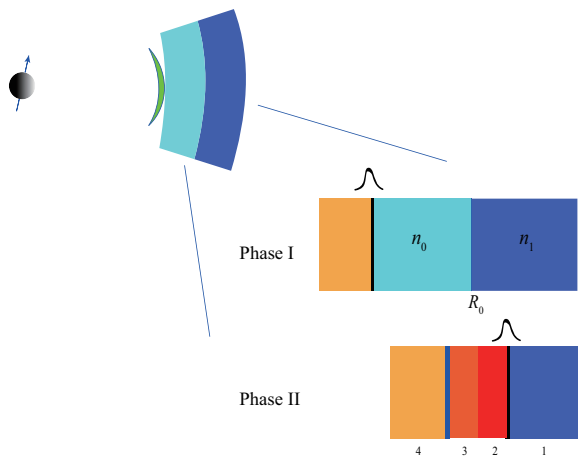}
\caption{The schematic picture of our density jump scenario. An ultra-relativistic ejecta from a magnetar flare collides with the leftover baryonic shell of previous flares. The shell is stratified and the outer part is denser ($n_0<n_1$). The deceleration of the ejecta will go through two phases and two FRB pulses are produced via synchrotron maser emission near the forward shock front (marked as the thick black line).}
\end{center}
\end{figure}

The matter ejected during magnetar flares usually consists of two components: a relativistic ejecta and a subrelativistic baryonic shell. The shell could possibly have some structure due to unsteady ejection. For instance, \citet{Beloborodov2020} suggested a distribution of ejected mass over velocity $m(v)\propto v^{\xi}$ with $\xi>0$, which implies an increasing density gradient with radius. In this work, the radial density profile is simplified as a jump between two constant densities, which is written as
\bea
n(r)=\left\{
\begin{array}{lr}
n_0, & \,\,\,\,r<R_0,  \\
n_1, & \,\,\,\,r\geqslant R_0, \\
\end{array}
\right.
\label{eq:density}
\ena
where $n_0<n_1$ is assumed. Similar to \citet{Metzger2019}, the relativistic ejecta collides with the baryonic shell characterized by Eq.(\ref{eq:density}) from the previous flare. This scenario is illustrated in Figure \ref{fig1}.

The dynamic evolution of the system goes through two phases. In Phase I, forward and reverse shocks are formed when the ejecta collides with the inner shell. The reverse shock crosses the ejecta in a time $t_{\rm cr, I}\simeq\delta t$ \citep{Sari1995}, where $\delta t$ is the flare duration. The first pulse of FRB 200428 is probably produced via synchrotron maser emission at the crossing radius, where both the luminosity and energy of the forward shock is at its maximum. After $t_{\rm cr, I}$, the evolution of the shocked region approaches the Blandford-McKee (BM) self-similar form \citep{Blandford1976}.  When the blast wave arrives at $R_0$, it feels stronger resistance from the denser medium and new forward, reverse shocks are formed. There are four separated regions: unshocked high-density outer shell, forward-shocked outer shell, reverse-shocked inner shell and unshocked hot inner shell, marked as 1, 2, 3, 4 respectively. The main difference with Phase I is that region 4 is hot, leading to the different equations of the jump conditions \citep{Zhang2002, Dai2002}. The second pulse is produced by the new forward shock near $R_0$ since its luminosity decreases with radius.

\section{Constraining density structure from double-peaked FRB pulses}
\label{sec3}

The energy and duration of the magnetar flare are denoted as $E$ and $\delta t$ respectively. In Phase I, the reverse shock crossing time is $t_{\rm cr, I}\simeq\delta t$. Before crossing, the Lorentz factor of the shocked region is $\Gamma_{\rm I}(t\leq t_{\rm cr, I})=(f_{\rm I}\Gamma_{\rm ej}^2/4)^{1/4}$ \citep{Sari1995}, where $f_{\rm I}\equiv E/(4\pi r^2m_p n_0c^3\delta t\Gamma_{\rm ej}^2)$ and  $\Gamma_{\rm ej}$ is the initial Lorentz factor of the ejecta. Since $t_{\rm cr, I}=R_{\rm cr, I}/(2\Gamma_{\rm cr, I}^2c)$, the crossing radius $R_{\rm cr, I}$ can be obtained,
\bea
R_{\rm cr, I}&=&\left(\frac{E\delta t}{4\pi m_pcn_0}\right)^{1/4}\nonumber\\
&=&3.55\times10^{11}E_{40}^{1/4}\delta t_{-3}^{1/4}n_{0,3}^{-1/4}\,\rm cm.
\ena
The Lorentz factor at crossing radius is then $\Gamma_{\rm cr, I}=(R_{\rm cr, I}/(2c\delta t))^{1/2}$. After crossing, the shock enters the BM self-similar evolution stage that implies $\Gamma\propto r^{-3/2}$. Thus, a smooth transition at $R_{\rm cr, I}$ indicates
\bea
\Gamma_{\rm I}(t> t_{\rm cr, I})=\Gamma_{\rm cr, I}(r/R_{\rm cr, I})^{-3/2}.
\label{eq:phaseI}
\ena

The BM evolution of Phase I suddenly ends as the blast wave encounters the density jump at $R_0$. From observation we have the time separation between two pulses is $\Delta T=28.91\,\rm ms$, thus we can determine $R_0$ by letting
\bea
\Delta T=\int_{R_{\rm cr,I}}^{R_0}{\frac{{\rm d}r}{2\Gamma_{\rm I}^2c}}.
\ena
Then we get
\bea
R_0=R_{\rm cr, I}\left(\frac{4\Delta T}{\delta t}+1\right)^{1/4}.
\label{eq:R0}
\ena
Since region 4 is hot, the jump conditions between region 3 and 4 are \citep{Zhang2002, Dai2002}
\bea
\left(\frac{n_3}{n_4}\right)^2&=&\frac{(e_3/e_4)(1+3e_3/e_4)}{3+e_3/e_4},\\
\gamma_{34}^2&=&\frac{(1+3e_3/e_4)(3+e_3/e_4)}{16e_3/e_4},
\ena
where $e,\,n$ are energy density, particle number density and $\gamma_{34}$ is the relative Lorentz factor between region 3 and 4. The initial Lorentz factor of region 4 is $\gamma_4=\Gamma_{\rm I}(R_0)$. Defining $f\equiv e_4/(n_1m_pc^2)$ and combing the density jump equations of region 1 and 2, the solution corresponding to a relativistic reverse shock is \citep{Dai2002}
\bea
\gamma_2&=&\gamma_3=\frac{\gamma_4^{1/2}f^{1/4}}{3^{1/4}},\label{eq:phaseII}\\
\gamma_{34}&=&\frac{3^{1/4}\gamma_4^{1/2}}{2f^{1/4}}\gg1. \label{eq:gam34}
\ena
The above solution requires the density jump ratio $n_1/n_0>64/3$ and this will be justified later for FRB 200428 case.

Now we can deduce the densities $n_0,\,n_1$ from pulse properties. From observation we know pulse I of FRB 200428 was detected only by CHIME and pulse II was detected simultaneously by CHIME and STARE2. The second pulse is found much more luminous than the first one \citep{CHIME2020b, Bochenek2020}. Here we scale the luminosity ratio as $\sim$10. Since the shock luminosity is $L_{\rm sh}\simeq4\pi r^2\Gamma^4nm_pc^3$, the peak FRB luminosity can be obtained as $\nu L_\nu|_{\rm pk}\simeq f_{\xi}L_{\rm sh}$, where $f_{\xi}$ is the maser efficiency. Then the luminosity ratio of two peaks is
\bea
\frac{\nu L_\nu|_{\rm pk, II}}{\nu L_\nu|_{\rm pk, I}}&\sim&\frac{f_{\xi_{\rm II}}4\pi R_0^2n_1\gamma_2(R_0)^4m_pc^3}{f_{\xi_{\rm I}}4\pi R_{\rm cr,I}^2n_0\Gamma_{\rm cr,I}^4m_pc^3}\nonumber\\
&\simeq&\frac{4}{3}\left(\frac{R_0}{R_{\rm cr,I}}\right)^{-4}\frac{f_{\xi_{\rm II}}}{f_{\xi_{\rm I}}},\quad\,
\ena
therefore
\bea
\frac{f_{\xi_{\rm II}}}{f_{\xi_{\rm I}}}=7.5\left(\frac{\nu L_\nu|_{\rm pk, II}}{\nu L_\nu|_{\rm pk, I}}\right)_1\left(\frac{4\Delta T}{\delta t}+1\right).
\label{eq:eff}
\ena
PIC simulations suggest that $f_\xi$ depends strongly on the upstream magnetization parameter $\sigma$, and $f_\xi\sim7\times10^{-4}/\sigma^2$ for $\sigma\geq1$ \citep{Plotnikov2019}. For typical flare duration $\delta t\sim1\,\rm ms$, Eq.(\ref{eq:eff}) implies that $f_{\xi_{\rm I}}\ll f_{\xi_{\rm II}}$, thus $\sigma_{\rm I}>1$ is always expected. For instance, if we assume the outer shell has $\sigma_{\rm II}\sim0.1$ that corresponds to a maximum maser efficiency of $f_{\xi_{\rm II}}\sim0.05$, then $f_{\xi_{\rm I}}\sim5.7\times10^{-5}$ and $\sigma_{\rm I}\sim3.5$. This makes sense because the inner shell should have higher magnetization \citep{Beloborodov2020}.

The peak frequency for synchrotron maser emission is $\nu_{\rm pk}=(3/2\pi)\Gamma\omega_{\rm p}\max[1,\sqrt{\sigma}]$ \citep{Plotnikov2019}, where $\omega_{\rm p}=(4\pi n_ee^2/m_e)^{1/2}$ is the plasma frequency. However, the observed FRB frequency is shifted to a higher value due to absorption of low-energy photons by induced Compton scattering (ICS). The optical depth of ICS near $\nu_{\rm pk}$ is approximated as \citep{Lyubarsky2008, Metzger2019}
\bea
\tau(\nu_{\rm pk})&\simeq&\frac{1}{10}\frac{3}{64\pi^2}\frac{\sigma_{\rm T}}{m_e}\frac{ctn}{r^2}\frac{\partial}{\partial\nu}\left(\frac{L_\nu}{\nu}\right)\bigg|_{\rm pk},\nonumber\\
&\simeq&\frac{3}{640\pi^2}\frac{\sigma_{\rm T}}{m_e}\frac{\nu L_\nu|_{\rm pk}ctn}{\nu_{\rm pk}^3r^2},
\label{eq:ICS}
\ena
where the Thompson cross section $\sigma_{\rm T}=8\pi e^4/(3m_e^2c^4)$. Substituting $\nu L_\nu|_{\rm pk}$ into Eq.(\ref{eq:ICS}), we obtain
\bea
\tau(\nu_{\rm pk})\simeq\frac{\pi^2m_p}{1620m_e}f_\xi\nu_{\rm pk}t.
\label{eq:ICS2}
\ena
The observed FRB frequency $\nu_{\max}$ is reached as $\tau\simeq3$. Since $\tau(\nu)\propto\nu^{-4}$, we get
\bea
\frac{\nu_{\rm max}}{\nu_{\rm pk}}\simeq7.81\left(\frac{\tau}{3}\right)^{-1/4}f_{\xi,-3}^{1/4}\left(\frac{\nu_{\rm pk}}{\rm GHz}\right)^{1/4}t_{-3}^{1/4},
\ena
where the peak frequencies of two pulses
\bea
\nu_{\rm pk, I}=3\Gamma_{\rm I}\omega_{\rm p, I}\sqrt{\sigma_{\rm I}}/2\pi, \quad \nu_{\rm pk, II}=3\gamma_2\omega_{\rm p, II}/2\pi.
\label{eq:nupk}
\ena
Here we implicitly take $\sigma_{\rm II}<1$. The reason is that the effective maser efficiency of pulse I should not be too small since observationally the flux ratio between radio and X-ray band is $F_{\rm radio}/F_{X}\sim10^{-6}-10^{-5}$ \citep{Mereghetti2020}. The observed emission frequency is higher for pulse II and according to the waterfall plots we normalize them as $\nu_{\rm max,I}=400\,\rm MHz$, $\nu_{\rm max,II}=1.4\,\rm GHz$. Thus
\bea
\nu_{\rm pk,I}&\simeq&0.092\left(\frac{\tau}{3}\right)^{1/5}f_{\xi_{\rm I},-3}^{-1/5}\left(\frac{\nu_{\rm max, I}}{400\,\rm MHz}\right)^{4/5}t_{{\rm I},-3}^{-1/5}\,\rm GHz,\nonumber\\
\nu_{\rm pk,II}&\simeq&0.253\left(\frac{\tau}{3}\right)^{1/5}f_{\xi_{\rm II},-3}^{-1/5}\left(\frac{\nu_{\rm max, II}}{1.4\,\rm GHz}\right)^{4/5}t_{{\rm II},-3}^{-1/5}\,\rm GHz.\quad\,\,
\label{eq:nupk2}
\ena
These two peaks are produced at time $t_{\rm I}\simeq\delta t$ and $t_{\rm II}\simeq\delta t+\Delta T$ respectively. Substituting into Eq.(\ref{eq:nupk2}) and combining Eq.(\ref{eq:phaseI})(\ref{eq:phaseII})(\ref{eq:nupk}), we can obtain $n_0,\,n_1$ as
\bea
n_0&=&3.23\times10^3E_{40}^{-1/3}\delta t_{-3}^{7/15}f_{\xi_{\rm I},-3}^{2/15} \,{\rm cm^{-3}},\label{eq:n0}\\
n_1&=&5.27\times10^6n_0^{-1/2}E_{40}^{-1/2}\delta t_{-3}^{7/10}f_{\xi_{\rm II},-3}^{-4/5}\nonumber\\
&\,&\times\left(\frac{4\Delta T}{\delta t}+1\right)^{3/2}\left(\frac{\Delta T}{\delta t}+1\right)^{-4/5}\,\rm cm^{-3}.\quad\,\,\label{eq:n1_1}
\ena
The above expression for $n_1$ is valid only if $\gamma_2>1$. In the situation of $\gamma_2\simeq1$, from Eq.(\ref{eq:nupk})(\ref{eq:nupk2}) we get
\bea
n_1=8.82\times10^7\delta t_{-3}^{-2/5}f_{\xi_{\rm II},-3}^{-2/5}\left(\frac{\Delta T}{\delta t}+1\right)^{-2/5}\,\rm cm^{-3}. \quad\label{eq:n1_2}
\ena

\begin{figure}
\label{fig2}
\begin{center}
\includegraphics[width=0.45\textwidth]{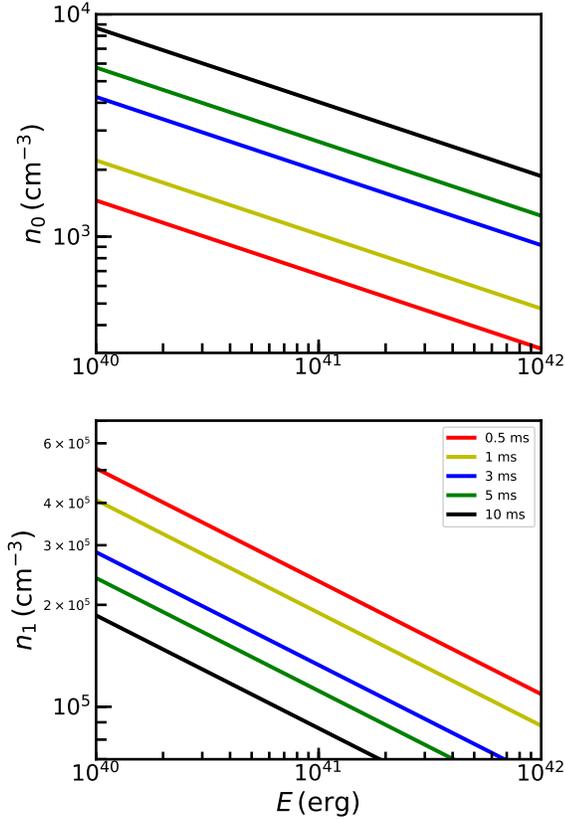}
\caption{The numerically-solved densities $n_0,\,n_1$ for different flare parameter sets $E,\,\delta t$. Different colored lines corresponds to different adopted $\delta t$, which is illustrated in the upper-right corner of lower panel. The behaviour of $n_0,\,n_1$ matches Eq.(\ref{eq:n0})(\ref{eq:n1_1}) well. The maser efficiency $f_{\xi_{\rm II}}=5\times10^{-2}$ is assumed and $f_{\xi_{\rm I}}$ is obtained according to Eq.(\ref{eq:eff}). For typical parameters of our interest, the requirement of a relativistic reverse shock is always fulfilled.}
\end{center}
\end{figure}

The dependences of $n_0,\,n_1$ on flare parameters $E,\,\delta t$ are shown in Figure \ref{fig2}. We adopt five typical values of $\delta t$ and the behavior of $n_0,\,n_1$ follows Eq.(\ref{eq:n0})(\ref{eq:n1_1}). From the figure we can see for the parameter space of our interest, $n_1/n_0>64/3$ is always ensured and a relativistic reverse shock in Phase II is expected. This can be understood from the ratio of peak frequencies of two pulses
\bea
\frac{\nu_{\rm pk,II}}{\nu_{\rm pk,I}}=\frac{1}{\sqrt{\sigma_{\rm I}}}\frac{\gamma_2n_1^{1/2}}{\Gamma_{\rm cr, I}n_0^{1/2}}<\frac{1}{\sqrt{\sigma_{\rm I}}}\frac{\gamma_4}{\Gamma_{\rm cr,I}}\left(\frac{n_1}{n_0}\right)^{1/2}.
\ena
From Eq.(\ref{eq:nupk2}) we know $\nu_{\rm pk,II}/{\nu_{\rm pk,I}}\sim$ a few, thus
\bea
\frac{n_1}{n_0}\gg\sigma_{\rm I}\left(\frac{4\Delta T}{\delta t}+1\right)^{3/4}
\ena
Usually with typical parameters $n_1/n_0>64/3$ is always satisfied. This validates the expressions of Eq.(\ref{eq:phaseII})(\ref{eq:gam34}).

The reverse shock crossing radius in Phase II can be evaluated as follows. The initial total particle number in region 4 is the swept-up inner shell before $R_0$, i.e., $N_4=4\pi R_0^3n_0 /3$. The total number in region 3 during crossing is $N_3(r)=\int_{R_0}^r{4\pi r^2n_4\gamma_{34}\beta_{34}/(\gamma_3\beta_3){\rm d}r}$ \citep{Dai2002}. During crossing, region 4 expands adiabatically so $n_4(r)=4\gamma_4n_0(r/R_0)^{-3}$. At the crossing radius $R_{\rm cr,II}$ we should have $N_3(R_{\rm cr,II})=N_4$. Substituting Eq.(\ref{eq:phaseII})(\ref{eq:gam34}) we obtain
\bea
R_{\rm cr,II}=R_0\left(1+\frac{2}{3^{3/2}}n_0^{1/2}n_1^{-1/2}\right)^{1/2}.
\ena
The crossing time is then
\bea
t_{\rm cr,II}=\int_{R_0}^{R_{\rm cr,II}}\frac{{\rm d}R}{2\gamma_2^2c}.
\ena
For the fiducial values $E=10^{40}\,\rm erg$, $\delta t=1\,\rm ms$, we get $R_0=9.58\times10^{11}\,\rm cm$, $R_{\rm cr,II}=9.71\times10^{11}\,\rm cm$ and $t_{\rm cr, II}=19.6\,{\rm ms}$. The total mass of the baryonic shell is $M=M_{\rm in}+M_{\rm out}\sim(4/3)\pi m_p[R_0^3n_0+(R_{\rm cr, II}^3-R_0^3)n_1]\simeq1.21\times10^{17}\,\rm g$, which is reasonably smaller than the ejected mass of 2004 giant flare from SGR 1806-20 \citep{Gelfand2005}. The mass ratio of two shells is $M_{\rm out}/M_{\rm in}\sim 8$, while the initial velocity ratio is approximated as $v_{\rm out}/v_{\rm in}\sim R_0/R_{\rm cr,I}\sim 3$. Therefore, our scenario is consistent with the power-law mass distribution $m(v)\propto v^{\xi}$ with $\xi\sim2$ if mass distribution is shaped by gravity \citep{Beloborodov2020}. Note that the mass ejection by magnetar flares is concentrated in short episodes. The ejection rate is not known well and the slower shell might be ejected earlier. In this case the faster shell would catch up and interact with the slower shell. The interaction of two shells could also lead to a density jump and this has been discussed both analytically and numerically in the context of stellar winds \citep{Luo1991,Dwarkadas1998,Ramirez-Ruiz2001}. However, a jump ratio of $> 20$ may be not very common since observationally the upward drifting of FRBs is rarer than downward drifting and till now only appeared for FRB 200428 and FRB 190611 \citep{Day2020}.

\section{The Associated X-ray burst}
\label{sec4}
For magnetized shocks, the electrons are likely to be heated to a thermal distribution, with a mean Lorentz factor $\bar{\gamma}\sim(1/2)(m_p/m_e)\Gamma$ \citep{Giannios2009}. Therefore in Phase I, the typical synchrotron frequency after crossing ($t\geq\delta t$) is
\bea
h\nu_{\rm syn, I}=h\frac{\bar{\gamma}^2eB}{2\pi m_ec}\Gamma_{\rm I}=5.94\,{\rm MeV}\,\sigma_{{\rm I},0}^{1/2}E_{40}^{1/2}t_{-3}^{-3/2},
\label{eq:synI}
\ena
where the post-shock magnetic field strength is $B=(64\pi\sigma_{\rm I}\Gamma^2m_pc^2n)^{1/2}$. The cooling frequency is
\bea
h\nu_{\rm c,I}&=&h\frac{\gamma_{\rm c}^2eB}{2\pi m_ec}\Gamma_{\rm I}=h\frac{eB}{2\pi m_ec}\Gamma_{\rm I}\left(\frac{6\pi m_ec}{\sigma_{\rm T}\Gamma_{\rm I}B^2t}\right)^2\nonumber\\
&=&11.69\,{\rm keV}\,\sigma_{{\rm I},0}^{-3/2}E_{40}^{-1/6}f_{\xi_{\rm I},-3}^{-2/15}\delta t_{-3}^{-7/15}t_{-3}^{-1/2}.
\label{eq:coolI}
\ena
Initially $\nu_{\rm syn, I}>\nu_{\rm c,I}$, so fast cooling is expected. Letting $\nu_{\rm syn, I}=\nu_{\rm c,I}$, the transition time to slow cooling is
\bea
t_{\rm c}\simeq 0.5\,{\rm s}\, \sigma_{{\rm I},0}^2E_{40}^{2/3}f_{\xi_{\rm I},-3}^{2/15}\delta t_{-3}^{7/15}.
\ena
The observing X-ray band is $\nu_X\sim20-200 \,\rm keV$ and from Eq.(\ref{eq:synI})(\ref{eq:coolI}) we know initially $\nu_{\rm c, I}<\nu_X<\nu_{\rm syn,I}$. The observed synchrotron luminosity is then $\nu L_{\nu,\rm syn}\propto(\nu_X/\nu_{\rm syn, I})^{1/2}$. As $\nu_{\rm syn, I}$ decreases with time, the X-ray luminosity increases until $\nu_{\rm syn, I}\simeq\nu_X$. The delay of X-ray peak arrival time with respect to radio peak \citep[$\sim$several ms,][]{Mereghetti2020} is expected. During phase II, regions 2, 3, and 4 are hot and their synchrotron emission can be important. For region 4, its emission is brightest at $R=R_0$ where $N_4$ is at its maximum. Since $\nu_{\rm syn,I}\propto \Gamma^4$, we have
\bea
\frac{h\nu_{{\rm syn},4}}{h\nu_{\rm syn, I}(t_{\rm cr, I})}=\frac{\gamma_4^4}{\Gamma_{\rm cr, I}^4}=\left(\frac{R_0}{R_{\rm cr, I}}\right)^{-6}.\,\,
\label{eq:nu4ratio}
\ena
Substituting Eq.(\ref{eq:R0})(\ref{eq:synI}) we obtain
\bea
h\nu_{{\rm syn},4}&=&h\nu_{\rm syn, I}(\delta t)\left(\frac{4\Delta T}{\delta t}+1\right)^{-3/2}.
\ena
Furthermore, the synchrotron frequencies of region 2 and 3 are respectively
\bea
\frac{h\nu_{{\rm syn},2}}{h\nu_{{\rm syn}, 4}}&=&\frac{\gamma_2^4\sqrt{\sigma_{\rm II}n_1}}{\gamma_4^4\sqrt{\sigma_{\rm I}n_0}}=\frac{4}{3}\sqrt{\frac{n_0\sigma_{\rm II}}{n_1\sigma_{\rm I}}}\left(\frac{R}{R_0}\right)^{-4},\nonumber\\
\frac{h\nu_{{\rm syn},3}}{h\nu_{{\rm syn}, 4}}&=&\frac{\gamma_3^4\sqrt{n_4}}{\gamma_4^4\sqrt{n_0}}=\frac{8\sqrt{\gamma_4}}{3}\frac{n_0}{n_1}\left(\frac{R}{R_0}\right)^{-11/2},
\label{eq:nu2ratio}
\ena
The emission from these two regions is brightest near $R_0$ since the shock luminosity $L_{\rm sh, II}\propto R^2\gamma_2^4\propto R^{-2}$, which means that the emission of region 2, 3, 4 reaches their maximum almost simultaneously in the local frame. However, in the observer frame they will be detected at different time. The emission of region 4 will arrive first since $\gamma_4>\gamma_2=\gamma_3$. The delay time of region 2 \& 3 is estimated as
\bea
\Delta t_{2,3}\sim\int_{R_0}^{R_{\rm cr, II}}{\frac{{\rm d}R}{2c}\left(\frac{1}{\gamma_2^2}-\frac{1}{\gamma_4^2}\right)}.
\ena
For typical values $E=10^{40}\,\rm erg$ and $\delta t=1\,\rm ms$ we get $\Delta t_{\rm 2,3}\sim 18\,\rm ms$, being marginally consistent with the time interval between the second and third peak in HXMT and INTEGRAL X-ray light curves \citep{Li2020,Mereghetti2020}. Moreover, it is possible to adjust parameters $E,\delta t, \sigma_{\rm I},\,\sigma_{\rm II}$ to make sure that $h\nu_{{\rm syn, I}},\,h\nu_{{\rm syn},2},\,h\nu_{{\rm syn},4}$ are in the observing band.

Nevertheless, there are two discrepancies with X-ray observations. First, from Eq.(\ref{eq:nu4ratio})(\ref{eq:nu2ratio}) we know $h\nu_{{\rm syn},2}<h\nu_{{\rm syn},4}<h\nu_{{\rm syn, I}}$. However, the peak energy of three peaks in INTEGRAL light curve seems increasing with time \citep{Mereghetti2020}. This problem will be alleviated if non-thermal electrons are present in the downstream and the superposition of two components could change the peak energy of the observed spectrum. On one way, the non-thermal component of the FRB-associated X-ray burst has been identified observationally \citep{Li2020}. On the other way, non-thermal acceleration of electrons by magnetized shocks in the presence of ions has been verified in PIC simulations \citep{Amato2006}. The second inconsistency is that the shock luminosity in Phase I is higher than that of Phase II, so the first X-ray peak is expected to be brighter than the second one. Observationally it is the second peak that has the highest luminosity. This contradiction can be reconciled if an efficiency parameter is introduced. PIC simulations show that the non-thermal acceleration efficiency varies with magnetization and number fraction of ions \citep{Amato2006}. Therefore, we should expect different X-ray radiation efficiencies of Phase I and II. The fitting of X-ray bursts properties needs additional free parameters thus is left for future work. Moreover, the observed X-ray flux should contain thermal emission from the magnetar and therefore can be merely considered as an upper limit for shock downstream emission.

\section{Discussion and Conclusions}
\label{sec5}
In this work, we have focused on the temporal behavior of FRB 200428 and the production of two pulses is ascribed to the shock propagation in a stratified medium. Since the second pulse has a higher frequency, a jump from low density shell to high density shell is expected. Both the luminosity and peak frequency of FRB emission can be well explained in the synchrotron maser scenario, assuming typical flare parameters. The density of two shells can be determined self-consistently. Synchrotron emission of different post-shock regions are in the X-ray band and can be responsible for a portion of FRB-associated X-ray burst fluence. Moreover, the arrival times of three peaks of the X-ray bursts could match the model prediction well.

Several previous works have discussed the application of synchrotron maser model to FRB 200428  \citep{Margalit2020,Yu2020,Wu2020}, however, none of them have explained how double peaks can be formed. On the contrary, \citet{Lu2020b} and \citet{WangJS2020} attributed these two peaks to two separated ejectas, and they put strong constraint on this scenario. Different from their assumption, only one ejecta exists in our scenario and it is the density jump that leads to two pulses. The recurrence of double-peaked character \citep{Kirsten2020} are in favor of our scenario, since one should not expect the magnetar produce multiple ejectas during each flare.

Alternatively, a few works have proposed that coherent curvature emission could be responsible for FRB 200428 \citep{Lu2020b,Yang2020b,WangWY2020,Dai2020b,Geng2020}. For this mechanism, coherence is achieved if the phases of EM waves emitted by each individual electron in the magnetosphere are near the same \citep{Ginzburg1975,Benford1977}. \citet{Lu2020b} and \citet{Yang2020b} assumed a disturbance from magnetar surface spreads and launches Alfv\'en waves in the polar region. In their scenario, the NS surface is heated precedently and X-rays should appear earlier than FRB pulses, since FRB is not yet produced until Alfv\'en waves reach the charge starvation radius. This inconsistence with observation also applies to the magnetar-asteroid model of \citet{Geng2020} but not to \citet{Dai2020b} since coherent curvature emission occurs before the asteroid matter is accreted onto the magnetar. However, the third X-ray peak in INTEGRAL and {\it Insight}-HXMT light curves are not expected in all these above models.

The structure of baryonic shell is expected to be ubiquitous since the mass outflow is unsteady during magnetar flares. This density jump scenario can potentially be applied to FRBs with multiple pulses once upgraded properly. We note that the pulse properties depend strongly on the external medium in synchrotron maser models \citep{Metzger2019, Beloborodov2020}. Therefore, different density structure can lead to a variety of FRB light curves. Other kinds of density structure can be considered for individual FRBs in the future, which is analogous to afterglow modeling in gamma-ray burst studies.
\acknowledgements

This work is supported by the National Key Research and Development Program of China (Grant No. 2017YFA0402600) and the National Natural Science Foundation of China (Grant No. 11833003, 11903018 and 11851305). DX is also supported by the Natural Science Foundation for the Youth of Jiangsu Province (Grant NO. BK20180324).

\bibliographystyle{aasjournal}
\bibliography{FRB}

\begin{thebibliography}{}
\expandafter\ifx\csname natexlab\endcsname\relax\def\natexlab#1{#1}\fi
\providecommand{\url}[1]{\href{#1}{#1}}
\providecommand{\dodoi}[1]{doi:~\href{http://doi.org/#1}{\nolinkurl{#1}}}
\providecommand{\doeprint}[1]{\href{http://ascl.net/#1}{\nolinkurl{http://ascl.net/#1}}}
\providecommand{\doarXiv}[1]{\href{https://arxiv.org/abs/#1}{\nolinkurl{https://arxiv.org/abs/#1}}}

\bibitem[{{Alsop} \& {Arons}(1988)}]{Alsop1988}
{Alsop}, D., \& {Arons}, J. 1988, Physics of Fluids, 31, 839,
  \dodoi{10.1063/1.866765}

\bibitem[{{Amato} \& {Arons}(2006)}]{Amato2006}
{Amato}, E., \& {Arons}, J. 2006, \apj, 653, 325, \dodoi{10.1086/508050}

\bibitem[{{Bannister} {et~al.}(2019){Bannister}, {Deller}, {Phillips},
  {Macquart}, {Prochaska}, {Tejos}, {Ryder}, {Sadler}, {Shannon}, {Simha},
  {Day}, {McQuinn}, {North-Hickey}, {Bhandari}, {Arcus}, {Bennert}, {Burchett},
  {Bouwhuis}, {Dodson}, {Ekers}, {Farah}, {Flynn}, {James}, {Kerr}, {Lenc},
  {Mahony}, {O'Meara}, {Os{\l}owski}, {Qiu}, {Treu}, {U}, {Bateman}, {Bock},
  {Bolton}, {Brown}, {Bunton}, {Chippendale}, {Cooray}, {Cornwell}, {Gupta},
  {Hayman}, {Kesteven}, {Koribalski}, {MacLeod}, {McClure-Griffiths},
  {Neuhold}, {Norris}, {Pilawa}, {Qiao}, {Reynolds}, {Roxby}, {Shimwell},
  {Voronkov}, \& {Wilson}}]{Bannister2019}
{Bannister}, K.~W., {Deller}, A.~T., {Phillips}, C., {et~al.} 2019, Science,
  365, 565, \dodoi{10.1126/science.aaw5903}

\bibitem[{{Beloborodov}(2017)}]{Beloborodov2017}
{Beloborodov}, A.~M. 2017, \apjl, 843, L26, \dodoi{10.3847/2041-8213/aa78f3}

\bibitem[{{Beloborodov}(2020)}]{Beloborodov2020}
---. 2020, \apj, 896, 142, \dodoi{10.3847/1538-4357/ab83eb}

\bibitem[{{Benford} \& {Buschauer}(1977)}]{Benford1977}
{Benford}, G., \& {Buschauer}, R. 1977, \mnras, 179, 189,
  \dodoi{10.1093/mnras/179.2.189}

\bibitem[{{Blandford} \& {McKee}(1976)}]{Blandford1976}
{Blandford}, R.~D., \& {McKee}, C.~F. 1976, Physics of Fluids, 19, 1130,
  \dodoi{10.1063/1.861619}

\bibitem[{{Bochenek} {et~al.}(2020){Bochenek}, {Ravi}, {Belov}, {Hallinan},
  {Kocz}, {Kulkarni}, \& {McKenna}}]{Bochenek2020}
{Bochenek}, C.~D., {Ravi}, V., {Belov}, K.~V., {et~al.} 2020, arXiv e-prints,
  arXiv:2005.10828.
\newblock \doarXiv{2005.10828}

\bibitem[{{CHIME/FRB Collaboration} {et~al.}(2019{\natexlab{a}}){CHIME/FRB
  Collaboration}, {Amiri}, {Bandura}, {Bhardwaj}, {Boubel}, {Boyce}, {Boyle},
  {. Brar}, {Burhanpurkar}, {Cassanelli}, {Chawla}, {Cliche}, {Cubranic},
  {Deng}, {Denman}, {Dobbs}, {Fandino}, {Fonseca}, {Gaensler}, {Gilbert},
  {Gill}, {Giri}, {Good}, {Halpern}, {Hanna}, {Hill}, {Hinshaw}, {H{\"o}fer},
  {Josephy}, {Kaspi}, {Landecker}, {Lang}, {Lin}, {Masui}, {Mckinven},
  {Mena-Parra}, {Merryfield}, {Michilli}, {Milutinovic}, {Moatti}, {Naidu},
  {Newburgh}, {Ng}, {Patel}, {Pen}, {Pinsonneault-Marotte}, {Pleunis},
  {Rafiei-Ravandi}, {Rahman}, {Ransom}, {Renard}, {Scholz}, {Shaw}, {Siegel},
  {Smith}, {Stairs}, {Tendulkar}, {Tretyakov}, {Vanderlinde}, \&
  {Yadav}}]{CHIME2019a}
{CHIME/FRB Collaboration}, {Amiri}, M., {Bandura}, K., {et~al.}
  2019{\natexlab{a}}, \nat, 566, 235, \dodoi{10.1038/s41586-018-0864-x}

\bibitem[{{CHIME/FRB Collaboration} {et~al.}(2019{\natexlab{b}}){CHIME/FRB
  Collaboration}, {Andersen}, {Bandura}, {Bhardwaj}, {Boubel}, {Boyce},
  {Boyle}, {Brar}, {Cassanelli}, {Chawla}, {Cubranic}, {Deng}, {Dobbs},
  {Fandino}, {Fonseca}, {Gaensler}, {Gilbert}, {Giri}, {Good}, {Halpern},
  {Hill}, {Hinshaw}, {H{\"o}fer}, {Josephy}, {Kaspi}, {Kothes}, {Landecker},
  {Lang}, {Li}, {Lin}, {Masui}, {Mena-Parra}, {Merryfield}, {Mckinven},
  {Michilli}, {Milutinovic}, {Naidu}, {Newburgh}, {Ng}, {Patel}, {Pen},
  {Pinsonneault-Marotte}, {Pleunis}, {Rafiei-Ravandi}, {Rahman}, {Ransom},
  {Renard}, {Scholz}, {Siegel}, {Singh}, {Smith}, {Stairs}, {Tendulkar},
  {Tretyakov}, {Vanderlinde}, {Yadav}, \& {Zwaniga}}]{CHIME2019b}
{CHIME/FRB Collaboration}, {Andersen}, B.~C., {Bandura}, K., {et~al.}
  2019{\natexlab{b}}, \apjl, 885, L24, \dodoi{10.3847/2041-8213/ab4a80}

\bibitem[{{CHIME/FRB Collaboration} {et~al.}(2020){CHIME/FRB Collaboration},
  {Amiri}, {Andersen}, {Band ura}, {Bhardwaj}, {Boyle}, {Brar}, {Chawla},
  {Chen}, {Cliche}, {Cubranic}, {Deng}, {Denman}, {Dobbs}, {Dong}, {Fand ino},
  {Fonseca}, {Gaensler}, {Giri}, {Good}, {Halpern}, {Hessels}, {Hill},
  {H{\"o}fer}, {Josephy}, {Kania}, {Karuppusamy}, {Kaspi}, {Keimpema},
  {Kirsten}, {Landecker}, {Lang}, {Leung}, {Li}, {Lin}, {Marcote}, {Masui},
  {McKinven}, {Mena-Parra}, {Merryfield}, {Michilli}, {Milutinovic},
  {Mirhosseini}, {Naidu}, {Newburgh}, {Ng}, {Nimmo}, {Paragi}, {Patel}, {Pen},
  {Pinsonneault-Marotte}, {Pleunis}, {Rafiei-Ravandi}, {Rahman}, {Ransom},
  {Renard}, {Sanghavi}, {Scholz}, {Shaw}, {Shin}, {Siegel}, {Singh}, {Smegal},
  {Smith}, {Stairs}, {Tendulkar}, {Tretyakov}, {Vanderlinde}, {Wang}, {Wang},
  {Wulf}, {Yadav}, \& {Zwaniga}}]{CHIME2020a}
{CHIME/FRB Collaboration}, {Amiri}, M., {Andersen}, B.~C., {et~al.} 2020, \nat,
  582, 351, \dodoi{10.1038/s41586-020-2398-2}

\bibitem[{{Chittidi} {et~al.}(2020){Chittidi}, {Simha}, {Mannings},
  {Prochaska}, {Rafelski}, {Neeleman}, {Macquart}, {Tejos}, {Jorgenson},
  {Ryder}, {Day}, {Marnoch}, {Bhandari}, {Deller}, {Qiu}, {Bannister},
  {Shannon}, \& {Heintz}}]{Chittidi2020}
{Chittidi}, J.~S., {Simha}, S., {Mannings}, A., {et~al.} 2020, arXiv e-prints,
  arXiv:2005.13158.
\newblock \doarXiv{2005.13158}

\bibitem[{{Cordes} \& {Chatterjee}(2019)}]{Cordes2019}
{Cordes}, J.~M., \& {Chatterjee}, S. 2019, \araa, 57, 417,
  \dodoi{10.1146/annurev-astro-091918-104501}

\bibitem[{{Cruces} {et~al.}(2020){Cruces}, {Spitler}, {Scholz}, {Lynch},
  {Seymour}, {Hessels}, {Gouiff{\`e}s}, {Hilmarsson}, {Kramer}, \&
  {Munjal}}]{Cruces2020}
{Cruces}, M., {Spitler}, L.~G., {Scholz}, P., {et~al.} 2020, arXiv e-prints,
  arXiv:2008.03461.
\newblock \doarXiv{2008.03461}

\bibitem[{{Dai}(2020)}]{Dai2020b}
{Dai}, Z.~G. 2020, \apjl, 897, L40, \dodoi{10.3847/2041-8213/aba11b}

\bibitem[{{Dai} \& {Lu}(2002)}]{Dai2002}
{Dai}, Z.~G., \& {Lu}, T. 2002, \apjl, 565, L87, \dodoi{10.1086/339418}

\bibitem[{{Day} {et~al.}(2020){Day}, {Deller}, {Shannon}, {Qiu}, {Bannister},
  {Bhandari}, {Ekers}, {Flynn}, {James}, {Macquart}, {Mahony}, {Phillips}, \&
  {Prochaska}}]{Day2020}
{Day}, C.~K., {Deller}, A.~T., {Shannon}, R.~M., {et~al.} 2020, \mnras,
  \dodoi{10.1093/mnras/staa2138}

\bibitem[{{Dwarkadas} \& {Balick}(1998)}]{Dwarkadas1998}
{Dwarkadas}, V.~V., \& {Balick}, B. 1998, \apj, 497, 267,
  \dodoi{10.1086/305464}

\bibitem[{{Fonseca} {et~al.}(2020){Fonseca}, {Andersen}, {Bhardwaj}, {Chawla},
  {Good}, {Josephy}, {Kaspi}, {Masui}, {Mckinven}, {Michilli}, {Pleunis},
  {Shin}, {Tendulkar}, {Bandura}, {Boyle}, {Brar}, {Cassanelli}, {Cubranic},
  {Dobbs}, {Dong}, {Gaensler}, {Hinshaw}, {Land ecker}, {Leung}, {Li}, {Lin},
  {Mena-Parra}, {Merryfield}, {Naidu}, {Ng}, {Patel}, {Pen}, {Rafiei-Ravandi},
  {Rahman}, {Ransom}, {Scholz}, {Smith}, {Stairs}, {Vanderlinde}, {Yadav}, \&
  {Zwaniga}}]{Fonseca2020}
{Fonseca}, E., {Andersen}, B.~C., {Bhardwaj}, M., {et~al.} 2020, \apjl, 891,
  L6, \dodoi{10.3847/2041-8213/ab7208}

\bibitem[{{Gelfand} {et~al.}(2005){Gelfand}, {Lyubarsky}, {Eichler},
  {Gaensler}, {Taylor}, {Granot}, {Newton-McGee}, {Ramirez-Ruiz},
  {Kouveliotou}, \& {Wijers}}]{Gelfand2005}
{Gelfand}, J.~D., {Lyubarsky}, Y.~E., {Eichler}, D., {et~al.} 2005, \apjl, 634,
  L89, \dodoi{10.1086/498643}

\bibitem[{{Geng} {et~al.}(2020){Geng}, {Li}, {Li}, {Xiong}, {Kuiper}, \&
  {Huang}}]{Geng2020}
{Geng}, J.-J., {Li}, B., {Li}, L.-B., {et~al.} 2020, \apjl, 898, L55,
  \dodoi{10.3847/2041-8213/aba83c}

\bibitem[{{Giannios} \& {Spitkovsky}(2009)}]{Giannios2009}
{Giannios}, D., \& {Spitkovsky}, A. 2009, \mnras, 400, 330,
  \dodoi{10.1111/j.1365-2966.2009.15454.x}

\bibitem[{{Ginzburg} \& {Zhelezniakov}(1975)}]{Ginzburg1975}
{Ginzburg}, V.~L., \& {Zhelezniakov}, V.~V. 1975, \araa, 13, 511,
  \dodoi{10.1146/annurev.aa.13.090175.002455}

\bibitem[{{Hoshino} \& {Arons}(1991)}]{Hoshino1991}
{Hoshino}, M., \& {Arons}, J. 1991, Physics of Fluids B, 3, 818,
  \dodoi{10.1063/1.859877}

\bibitem[{{Hoshino} {et~al.}(1992){Hoshino}, {Arons}, {Gallant}, \&
  {Langdon}}]{Hoshino1992}
{Hoshino}, M., {Arons}, J., {Gallant}, Y.~A., \& {Langdon}, A.~B. 1992, \apj,
  390, 454, \dodoi{10.1086/171296}

\bibitem[{{Israel} {et~al.}(2016){Israel}, {Esposito}, {Rea}, {Coti Zelati},
  {Tiengo}, {Campana}, {Mereghetti}, {Rodriguez Castillo}, {G{\"o}tz},
  {Burgay}, {Possenti}, {Zane}, {Turolla}, {Perna}, {Cannizzaro}, \&
  {Pons}}]{Israel2016}
{Israel}, G.~L., {Esposito}, P., {Rea}, N., {et~al.} 2016, \mnras, 457, 3448,
  \dodoi{10.1093/mnras/stw008}

\bibitem[{{Katz}(2018)}]{Katz2018}
{Katz}, J.~I. 2018, Progress in Particle and Nuclear Physics, 103, 1,
  \dodoi{10.1016/j.ppnp.2018.07.001}

\bibitem[{{Kirsten} {et~al.}(2020){Kirsten}, {Snelders}, {Jenkins}, {Nimmo},
  {van den Eijnden}, {Hessels}, {Gawronski}, \& {Yang}}]{Kirsten2020}
{Kirsten}, F., {Snelders}, M., {Jenkins}, M., {et~al.} 2020, arXiv e-prints,
  arXiv:2007.05101.
\newblock \doarXiv{2007.05101}

\bibitem[{{Kothes} {et~al.}(2018){Kothes}, {Sun}, {Gaensler}, \&
  {Reich}}]{Kothes2018}
{Kothes}, R., {Sun}, X., {Gaensler}, B., \& {Reich}, W. 2018, \apj, 852, 54,
  \dodoi{10.3847/1538-4357/aa9e89}

\bibitem[{{Li} {et~al.}(2020){Li}, {Lin}, {Xiong}, {Ge}, {Li}, {Li}, {Lu},
  {Zhang}, {Tuo}, {Nang}, {Zhang}, {Xiao}, {Chen}, {Song}, {Xu}, {Liu}, {Jia},
  {Cao}, {Zhang}, {Qu}, {Liao}, {Zhao}, {Tan}, {Nie}, {Zhao}, {Zheng}, {Zheng},
  {Luo}, {Cai}, {Li}, {Xue}, {Bu}, {Chang}, {Chen}, {Chen}, {Chen}, {Chen},
  {Chen}, {Cui}, {Cui}, {Deng}, {Dong}, {Du}, {Fu}, {Gao}, {Gao}, {Gao}, {Gu},
  {Guan}, {Guo}, {Han}, {Huang}, {Huo}, {Jiang}, {Jiang}, {Jin}, {Jin}, {Kong},
  {Li}, {Li}, {Li}, {Li}, {Li}, {Li}, {Li}, {Liang}, {Liu}, {Liu}, {Liu},
  {Liu}, {Liu}, {Lu}, {Lu}, {Luo}, {Ma}, {Meng}, {Ou}, {Sai}, {Shang}, {Song},
  {Sun}, {Tao}, {Wang}, {Wang}, {Wang}, {Wang}, {Wang}, {Wen}, {Wu}, {Wu},
  {Wu}, {Xiao}, {Yang}, {Yang}, {Yang}, {Yang}, {Yi}, {Yin}, {You}, {Zhang},
  {Zhang}, {Zhang}, {Zhang}, {Zhang}, {Zhang}, {Zhang}, {Zhang}, {Zhang},
  {Zhang}, {Zhang}, {Zhang}, {Zhang}, {Zhang}, {Zhang}, {Zhang}, {Zhou},
  {Zhou}, {Zhu}, {Zhu}, \& {Zhuang}}]{Li2020}
{Li}, C.~K., {Lin}, L., {Xiong}, S.~L., {et~al.} 2020, arXiv e-prints,
  arXiv:2005.11071.
\newblock \doarXiv{2005.11071}

\bibitem[{{Lorimer} {et~al.}(2007){Lorimer}, {Bailes}, {McLaughlin},
  {Narkevic}, \& {Crawford}}]{Lorimer2007}
{Lorimer}, D.~R., {Bailes}, M., {McLaughlin}, M.~A., {Narkevic}, D.~J., \&
  {Crawford}, F. 2007, Science, 318, 777, \dodoi{10.1126/science.1147532}

\bibitem[{{Lu} {et~al.}(2020){Lu}, {Kumar}, \& {Zhang}}]{Lu2020b}
{Lu}, W., {Kumar}, P., \& {Zhang}, B. 2020, arXiv e-prints, arXiv:2005.06736.
\newblock \doarXiv{2005.06736}

\bibitem[{{Luo} \& {McCray}(1991)}]{Luo1991}
{Luo}, D., \& {McCray}, R. 1991, \apj, 379, 659, \dodoi{10.1086/170539}

\bibitem[{{Lyubarsky}(2008)}]{Lyubarsky2008}
{Lyubarsky}, Y. 2008, \apj, 682, 1443, \dodoi{10.1086/589435}

\bibitem[{{Lyubarsky}(2014)}]{Lyubarsky2014}
---. 2014, \mnras, 442, L9, \dodoi{10.1093/mnrasl/slu046}

\bibitem[{{Macquart} {et~al.}(2020){Macquart}, {Prochaska}, {McQuinn},
  {Bannister}, {Bhandari}, {Day}, {Deller}, {Ekers}, {James}, {Marnoch},
  {Os{\l}owski}, {Phillips}, {Ryder}, {Scott}, {Shannon}, \&
  {Tejos}}]{Macquart2020}
{Macquart}, J.~P., {Prochaska}, J.~X., {McQuinn}, M., {et~al.} 2020, \nat, 581,
  391, \dodoi{10.1038/s41586-020-2300-2}

\bibitem[{{Marcote} {et~al.}(2020){Marcote}, {Nimmo}, {Hessels}, {Tendulkar},
  {Bassa}, {Paragi}, {Keimpema}, {Bhardwaj}, {Karuppusamy}, {Kaspi}, {Law},
  {Michilli}, {Aggarwal}, {Andersen}, {Archibald}, {Bandura}, {Bower}, {Boyle},
  {Brar}, {Burke-Spolaor}, {Butler}, {Cassanelli}, {Chawla}, {Demorest},
  {Dobbs}, {Fonseca}, {Giri}, {Good}, {Gourdji}, {Josephy}, {Kirichenko},
  {Kirsten}, {Landecker}, {Lang}, {Lazio}, {Li}, {Lin}, {Linford}, {Masui},
  {Mena-Parra}, {Naidu}, {Ng}, {Patel}, {Pen}, {Pleunis}, {Rafiei-Ravandi},
  {Rahman}, {Renard}, {Scholz}, {Siegel}, {Smith}, {Stairs}, {Vanderlinde}, \&
  {Zwaniga}}]{Marcote2020}
{Marcote}, B., {Nimmo}, K., {Hessels}, J.~W.~T., {et~al.} 2020, \nat, 577, 190,
  \dodoi{10.1038/s41586-019-1866-z}

\bibitem[{{Margalit} {et~al.}(2020){Margalit}, {Beniamini}, {Sridhar}, \&
  {Metzger}}]{Margalit2020}
{Margalit}, B., {Beniamini}, P., {Sridhar}, N., \& {Metzger}, B.~D. 2020,
  \apjl, 899, L27, \dodoi{10.3847/2041-8213/abac57}

\bibitem[{{Melrose}(2017)}]{Melrose2017}
{Melrose}, D.~B. 2017, Reviews of Modern Plasma Physics, 1, 5,
  \dodoi{10.1007/s41614-017-0007-0}

\bibitem[{{Mereghetti} {et~al.}(2020){Mereghetti}, {Savchenko}, {Ferrigno},
  {G{\"o}tz}, {Rigoselli}, {Tiengo}, {Bazzano}, {Bozzo}, {Coleiro},
  {Courvoisier}, {Doyle}, {Goldwurm}, {Hanlon}, {Jourdain}, {Kienlin},
  {Lutovinov}, {Martin-Carrillo}, {Molkov}, {Natalucci}, {Onori}, {Panessa},
  {Rodi}, {Rodriguez}, {S{\'a}nchez-Fern{\'a}ndez}, {Sunyaev}, \&
  {Ubertini}}]{Mereghetti2020}
{Mereghetti}, S., {Savchenko}, V., {Ferrigno}, C., {et~al.} 2020, \apjl, 898,
  L29, \dodoi{10.3847/2041-8213/aba2cf}

\bibitem[{{Metzger} {et~al.}(2019){Metzger}, {Margalit}, \&
  {Sironi}}]{Metzger2019}
{Metzger}, B.~D., {Margalit}, B., \& {Sironi}, L. 2019, \mnras, 485, 4091,
  \dodoi{10.1093/mnras/stz700}

\bibitem[{{Murase} {et~al.}(2016){Murase}, {Kashiyama}, \&
  {M{\'e}sz{\'a}ros}}]{Murase2016}
{Murase}, K., {Kashiyama}, K., \& {M{\'e}sz{\'a}ros}, P. 2016, \mnras, 461,
  1498, \dodoi{10.1093/mnras/stw1328}

\bibitem[{{Petroff} {et~al.}(2019){Petroff}, {Hessels}, \&
  {Lorimer}}]{Petroff2019}
{Petroff}, E., {Hessels}, J.~W.~T., \& {Lorimer}, D.~R. 2019, \aapr, 27, 4,
  \dodoi{10.1007/s00159-019-0116-6}

\bibitem[{{Petroff} {et~al.}(2016){Petroff}, {Barr}, {Jameson}, {Keane},
  {Bailes}, {Kramer}, {Morello}, {Tabbara}, \& {van Straten}}]{Petroff2016}
{Petroff}, E., {Barr}, E.~D., {Jameson}, A., {et~al.} 2016, \pasa, 33, e045,
  \dodoi{10.1017/pasa.2016.35}

\bibitem[{{Platts} {et~al.}(2019){Platts}, {Weltman}, {Walters}, {Tendulkar},
  {Gordin}, \& {Kandhai}}]{Platts2019}
{Platts}, E., {Weltman}, A., {Walters}, A., {et~al.} 2019, \physrep, 821, 1,
  \dodoi{10.1016/j.physrep.2019.06.003}

\bibitem[{{Plotnikov} \& {Sironi}(2019)}]{Plotnikov2019}
{Plotnikov}, I., \& {Sironi}, L. 2019, \mnras, 485, 3816,
  \dodoi{10.1093/mnras/stz640}

\bibitem[{{Popov} \& {Postnov}(2013)}]{Popov2013}
{Popov}, S.~B., \& {Postnov}, K.~A. 2013, arXiv e-prints, arXiv:1307.4924.
\newblock \doarXiv{1307.4924}

\bibitem[{{Popov} {et~al.}(2018){Popov}, {Postnov}, \& {Pshirkov}}]{Popov2018}
{Popov}, S.~B., {Postnov}, K.~A., \& {Pshirkov}, M.~S. 2018, Physics Uspekhi,
  61, 965, \dodoi{10.3367/UFNe.2018.03.038313}

\bibitem[{{Prochaska} {et~al.}(2019){Prochaska}, {Macquart}, {McQuinn},
  {Simha}, {Shannon}, {Day}, {Marnoch}, {Ryder}, {Deller}, {Bannister},
  {Bhandari}, {Bordoloi}, {Bunton}, {Cho}, {Flynn}, {Mahony}, {Phillips},
  {Qiu}, \& {Tejos}}]{Prochaska2019}
{Prochaska}, J.~X., {Macquart}, J.-P., {McQuinn}, M., {et~al.} 2019, Science,
  366, 231, \dodoi{10.1126/science.aay0073}

\bibitem[{{Rajwade} {et~al.}(2020){Rajwade}, {Mickaliger}, {Stappers},
  {Morello}, {Agarwal}, {Bassa}, {Breton}, {Caleb}, {Karastergiou}, {Keane}, \&
  {Lorimer}}]{Rajwade2020b}
{Rajwade}, K.~M., {Mickaliger}, M.~B., {Stappers}, B.~W., {et~al.} 2020,
  \mnras, 495, 3551, \dodoi{10.1093/mnras/staa1237}

\bibitem[{{Ramirez-Ruiz} {et~al.}(2001){Ramirez-Ruiz}, {Dray}, {Madau}, \&
  {Tout}}]{Ramirez-Ruiz2001}
{Ramirez-Ruiz}, E., {Dray}, L.~M., {Madau}, P., \& {Tout}, C.~A. 2001, \mnras,
  327, 829, \dodoi{10.1046/j.1365-8711.2001.04762.x}

\bibitem[{{Ravi} {et~al.}(2019){Ravi}, {Catha}, {D'Addario}, {Djorgovski},
  {Hallinan}, {Hobbs}, {Kocz}, {Kulkarni}, {Shi}, {Vedantham}, {Weinreb}, \&
  {Woody}}]{Ravi2019}
{Ravi}, V., {Catha}, M., {D'Addario}, L., {et~al.} 2019, \nat, 572, 352,
  \dodoi{10.1038/s41586-019-1389-7}

\bibitem[{{Ridnaia} {et~al.}(2020){Ridnaia}, {Svinkin}, {Frederiks}, {Bykov},
  {Popov}, {Aptekar}, {Golenetskii}, {Lysenko}, {Tsvetkova}, {Ulanov}, \&
  {Cline}}]{Ridnaia2020}
{Ridnaia}, A., {Svinkin}, D., {Frederiks}, D., {et~al.} 2020, arXiv e-prints,
  arXiv:2005.11178.
\newblock \doarXiv{2005.11178}

\bibitem[{{Sari} \& {Piran}(1995)}]{Sari1995}
{Sari}, R., \& {Piran}, T. 1995, \apjl, 455, L143, \dodoi{10.1086/309835}

\bibitem[{{Tavani} {et~al.}(2020){Tavani}, {Casentini}, {Ursi}, {Verrecchia},
  {Addis}, {Antonelli}, {Argan}, {Barbiellini}, {Baroncelli}, {Bernardi},
  {Bianchi}, {Bulgarelli}, {Caraveo}, {Cardillo}, {Cattaneo}, {Chen}, {Costa},
  {Del Monte}, {Di Cocco}, {Di Persio}, {Donnarumma}, {Evangelista}, {Feroci},
  {Ferrari}, {Fioretti}, {Fuschino}, {Galli}, {Gianotti}, {Giuliani},
  {Labanti}, {Lazzarotto}, {Lipari}, {Longo}, {Lucarelli}, {Magro},
  {Marisaldi}, {Mereghetti}, {Morelli}, {Morselli}, {Naldi}, {Pacciani},
  {Parmiggiani}, {Paoletti}, {Pellizzoni}, {Perri}, {Perotti}, {Piano},
  {Picozza}, {Pilia}, {Pittori}, {Puccetti}, {Pupillo}, {Rapisarda},
  {Rappoldi}, {Rubini}, {Setti}, {Soffitta}, {Trifoglio}, {Trois},
  {Vercellone}, {Vittorini}, {Giommi}, \& {D' Amico}}]{Tavani2020}
{Tavani}, M., {Casentini}, C., {Ursi}, A., {et~al.} 2020, arXiv e-prints,
  arXiv:2005.12164.
\newblock \doarXiv{2005.12164}

\bibitem[{{Tendulkar} {et~al.}(2017){Tendulkar}, {Bassa}, {Cordes}, {Bower},
  {Law}, {Chatterjee}, {Adams}, {Bogdanov}, {Burke-Spolaor}, {Butler},
  {Demorest}, {Hessels}, {Kaspi}, {Lazio}, {Maddox}, {Marcote}, {McLaughlin},
  {Paragi}, {Ransom}, {Scholz}, {Seymour}, {Spitler}, {van Langevelde}, \&
  {Wharton}}]{Tendulkar2017}
{Tendulkar}, S.~P., {Bassa}, C.~G., {Cordes}, J.~M., {et~al.} 2017, \apjl, 834,
  L7, \dodoi{10.3847/2041-8213/834/2/L7}

\bibitem[{{The CHIME/FRB Collaboration} {et~al.}(2020){The CHIME/FRB
  Collaboration}, {:}, {Andersen}, {Band ura}, {Bhardwaj}, {Bij}, {Boyce},
  {Boyle}, {Brar}, {Cassanelli}, {Chawla}, {Chen}, {Cliche}, {Cook},
  {Cubranic}, {Curtin}, {Denman}, {Dobbs}, {Dong}, {Fandino}, {Fonseca},
  {Gaensler}, {Giri}, {Good}, {Halpern}, {Hill}, {Hinshaw}, {H{\"o}fer},
  {Josephy}, {Kania}, {Kaspi}, {Landecker}, {Leung}, {Li}, {Lin}, {Masui},
  {Mckinven}, {Mena-Parra}, {Merryfield}, {Meyers}, {Michilli}, {Milutinovic},
  {Mirhosseini}, {M{\"u}nchmeyer}, {Naidu}, {Newburgh}, {Ng}, {Patel}, {Pen},
  {Pinsonneault-Marotte}, {Pleunis}, {Quine}, {Rafiei-Ravandi}, {Rahman},
  {Ransom}, {Renard}, {Sanghavi}, {Scholz}, {Shaw}, {Shin}, {Siegel}, {Singh},
  {Smegal}, {Smith}, {Stairs}, {Tan}, {Tendulkar}, {Tretyakov}, {Vanderlinde},
  {Wang}, {Wulf}, \& {Zwaniga}}]{CHIME2020b}
{The CHIME/FRB Collaboration}, {:}, {Andersen}, B.~C., {et~al.} 2020, arXiv
  e-prints, arXiv:2005.10324.
\newblock \doarXiv{2005.10324}

\bibitem[{{Wang}(2020)}]{WangJS2020}
{Wang}, J.-S. 2020, \apj, 900, 172, \dodoi{10.3847/1538-4357/aba955}

\bibitem[{{Wang} {et~al.}(2020){Wang}, {Xu}, \& {Chen}}]{WangWY2020}
{Wang}, W.-Y., {Xu}, R., \& {Chen}, X. 2020, \apj, 899, 109,
  \dodoi{10.3847/1538-4357/aba268}

\bibitem[{{Waxman}(2017)}]{Waxman2017}
{Waxman}, E. 2017, \apj, 842, 34, \dodoi{10.3847/1538-4357/aa713e}

\bibitem[{{Wu} {et~al.}(2020){Wu}, {Zhang}, {Wang}, \& {Dai}}]{Wu2020}
{Wu}, Q., {Zhang}, G.~Q., {Wang}, F.~Y., \& {Dai}, Z.~G. 2020, arXiv e-prints,
  arXiv:2008.05635.
\newblock \doarXiv{2008.05635}

\bibitem[{{Yang} {et~al.}(2020){Yang}, {Zhu}, {Zhang}, \& {Wu}}]{Yang2020b}
{Yang}, Y.-P., {Zhu}, J.-P., {Zhang}, B., \& {Wu}, X.-F. 2020, arXiv e-prints,
  arXiv:2006.03270.
\newblock \doarXiv{2006.03270}

\bibitem[{{Yu} {et~al.}(2020){Yu}, {Zou}, {Dai}, \& {Yu}}]{Yu2020}
{Yu}, Y.-W., {Zou}, Y.-C., {Dai}, Z.-G., \& {Yu}, W.-F. 2020, arXiv e-prints,
  arXiv:2006.00484.
\newblock \doarXiv{2006.00484}

\bibitem[{{Zhang} \& {M{\'e}sz{\'a}ros}(2002)}]{Zhang2002}
{Zhang}, B., \& {M{\'e}sz{\'a}ros}, P. 2002, \apj, 566, 712,
  \dodoi{10.1086/338247}

\bibitem[{{Zhong} {et~al.}(2020){Zhong}, {Dai}, {Zhang}, \& {Deng}}]{Zhong2020}
{Zhong}, S.-Q., {Dai}, Z.-G., {Zhang}, H.-M., \& {Deng}, C.-M. 2020, \apjl,
  898, L5, \dodoi{10.3847/2041-8213/aba262}

\bibitem[{{Zhou} {et~al.}(2020){Zhou}, {Zhou}, {Chen}, {Wang}, {Vink}, \&
  {Wang}}]{Zhou2020}
{Zhou}, P., {Zhou}, X., {Chen}, Y., {et~al.} 2020, arXiv e-prints,
  arXiv:2005.03517.
\newblock \doarXiv{2005.03517}

\end{thebibliography}

\end{document}